\documentclass[dvipdfmx,color,twocolumn,preprintnumbers,amsmath,amssymb,superscriptaddress]{revtex4-1}

\usepackage[dvipdfmx]{graphicx}
\usepackage{dcolumn}
\usepackage[dvipdfmx]{color}

\usepackage{amsmath,bm}

\begin{document}

\preprint{YITP-17-53}

\title{Constraining the $\bar{K}N$ interaction from the $1S$ level shift of kaonic deuterium}

\author{Tsubasa Hoshino}
\affiliation{Department of Physics, Hokkaido University, Sapporo 060-0810, Japan}

\author{Shota Ohnishi}
\affiliation{Department of Physics, Hokkaido University, Sapporo 060-0810, Japan}

\author{Wataru Horiuchi}
\affiliation{Department of Physics, Hokkaido University, Sapporo 060-0810, Japan}

\author{Tetsuo Hyodo}
\affiliation{Yukawa Institute for Theoretical Physics, Kyoto University, Kyoto 606-8502, Japan}

\author{Wolfram Weise}
\affiliation{Yukawa Institute for Theoretical Physics, Kyoto University, Kyoto 606-8502, Japan}
\affiliation{Physik-Department, Technische Universit\"{a}t M\"{u}nchen, 85748 Garching, Germany}

\begin{abstract}
  Motivated by the precise measurement of the 1$S$ level shift of
  kaonic hydrogen,
  we perform accurate three-body calculations
for the spectrum of kaonic deuterium using a realistic antikaon-nucleon ($\bar{K}N$) interaction.
In order to describe both short- and long-range behavior of the kaonic atomic states,
we solve the three-body Schr\"odinger equation with
a superposition of a large number of correlated Gaussian basis functions
covering distances up to several hundreds of fm.
Transition energies between 1$S$, 2$P$ and 2$S$ states are determined with high precision.
The complex energy shift of the $1S$ level  of kaonic deuterium 
is found to be $\Delta E-i\Gamma/2= (670-i\,508)$ eV. 
The sensitivity of this level shift with respect to the isospin $I=1$ component of the $\bar{K}N$ interaction is examined.
It is pointed out that an experimental determination of the kaonic deuterium level shift within an uncertainty of 25 \% will provide a constraint for the $I=1$ component of the $\bar{K}N$ interaction significantly stronger
than that from kaonic hydrogen.
\end{abstract}

\maketitle

\section{INTRODUCTION}

In recent years systems involving antikaons ($\bar{K}=K^-, \bar{K}^0$) have been widely explored
in hadron and strangeness nuclear physics. Studies of the $\Lambda$(1405), as a $\bar{K}N$ quasibound state coupled to the $\pi\Sigma$ continuum, suggested early on~\cite{Dalitz:1959dn,Dalitz:1960du} that
the low-energy interaction between $\bar{K}$ and nucleon ($N$)
is strongly attractive. This implies the principal possibility of forming $\bar{K}$-nuclear quasibound states, or kaonic nuclei~\cite{ay2002,Yamazaki:2002uh,ay2007,Dote:2002db,Dote:2003ac}. Many experiments and theoretical calculations have been devoted to search for such exotic nuclear systems (see, for example, Refs.~\cite{Gal:2013vx,Gal:2016boi} for recent reviews). 

The $\bar{K}NN$ three-body system, as the lightest prototype of a kaonic nucleus, has been studied actively. A number of theoretical works suggested the existence of this quasibound state~\cite{Yamazaki:2002uh,ay2007,shev1,shev2,ikeda1,dote1,dote2,Wycech:2008wf,ikeda2,barn,dote3,Ohnishi17} and pointed to its possible signatures in production reactions~\cite{Koike:2009mx,YamagataSekihara:2008ji,Ohnishi:2013rix,Sekihara:2016vyd}, but no consensus has been reached so far; the quantitative results depend strongly on the $\bar{K}N$ interaction employed in the calculations.

 Experimentally, the existence of kaonic nuclei is controversial as well.
 In some reports a peak structure is observed around $100$ MeV below the
 $\bar{K}NN$ threshold~\cite{evi1,evi2,Ichikawa:2014ydh}. But, if interpreted as a $\bar{K}$-nuclear bound state,
 such a large binding energy could not be accounted for in any of the theoretical
 studies~\cite{Yamazaki:2002uh,ay2007,shev1,shev2,ikeda1,dote1,dote2,Wycech:2008wf,ikeda2,barn,dote3,Ohnishi17}.
 On the other hand, measurements reported in Refs.~\cite{leps,hades,jparc} did not find a corresponding prominent signal in their spectra. Recently the J-PARC E15 experiment~\cite{jparc2} observed a peak structure near the $\bar{K}NN$ threshold in the $^3$He$(K^-,\Lambda p)n$ reaction, which awaits further analysis and interpretation.

In the theoretical calculations of kaonic nuclei the
$\bar{K}N$ interaction below the $\bar{K}N$ threshold energy is an essential ingredient.
Since the subthreshold energy region cannot be directly accessed by $\bar{K}N$ scattering experiments,
extrapolations are necessary in order to construct the scattering amplitude
below the $\bar{K}N$ threshold. Such extrapolations are subject to uncertainties
in the $\bar{K}$$N$ interaction itself.

A kaonic atom in which a $K^-$ is bound to 
an ordinary nucleus by the Coulomb force,
is a useful object for investigating the $\bar{K}N$ interaction
just below threshold.
In an ordinary atomic systems electrons are bound exclusively by the Coulomb interaction.
In a kaonic atom the binding energy is determined by both Coulomb and strong $\bar{K}N$ interactions.
The purely Coulombic energy of the $1S$ atomic orbit,
$E_{1S}^{C}<0$, is shifted  to a complex energy $E_{1S}$
by the $\bar{K}N$ interaction and the absorptive transitions to lower energy $\pi\Lambda$ and  $\pi\Sigma$ channels. 
This $1S$ level shift and width, $\Delta E-i\Gamma/2=E_{1S}-E^{C}_{1S}$, reflects the $\bar{K}N$ interaction at threshold. Actually this level shift is measured by
the X-ray transition energy from $2P$ to $1S$, assuming that
the $2P$ state is not affected by the $\bar{K}N$ interaction. A prominent example is
the $1S$ level shift of kaonic hydrogen, the $K^-$-proton ($p$) atomic system, determined
as $\Delta E =283 \pm 36 \pm 6 $ eV and $\Gamma = 541 \pm 89 \pm 22 $ eV
by the SIDDHARTA experiment~\cite{sidd,sidd2}.

The SIDDHARTA data helped reducing significantly
the theoretical uncertainties of the
scattering amplitude at and below the $\bar{K}N$ threshold~\cite{ihw1,ihw2}. However, these
data are not sufficient in order to determine
the full isospin dependence of the $\bar{K}N$ interaction.
The isospin $I=0$ component is well constrained by
the properties of the $\Lambda$(1405) as a $\bar{K}N (I=0)$ state coupled to the strongly interacting $\pi\Sigma$ continuum.
Kaonic hydrogen involves both isospin $I=0$ and $I=1$ components equally, but a
large uncertainty still remains in
the $I=1$ component of the $\bar{K}N$ interaction~\cite{Kamiya:2016jqc}.

In order to extract further information on this $I=1$ component
we consider kaonic deuterium, the $K^{-}$-deuteron ($d$) atomic system.
In $K^{-}d$ the ratio of $I=0$ to $I=1$ is $1:3$,
therefore kaonic deuterium is expected to be a more sensitive probe than kaonic hydrogen for the $I=1$ component of the $\bar{K}N$ interaction. Its full isospin dependence can thus be determined more precisely
by combining data from kaonic hydrogen and deuterium.

In the SIDDHARTA experiment~\cite{sidd},
a kaonic deuterium run was performed simultaneously with the hydrogen run
but the statistics was not sufficient for a reliable separation of signal versus background. 
New experiments are now being planned to measure kaonic deuterium: J-PARC E57~\cite{Zmeskal:2015efj} and SIDDHARTA-2~\cite{sidd-2,Iliescu:2016zgx}.
The tentatively expected precision of these measurements is 60 eV (30 eV) for the shift and 140 eV (70 eV) for the width at J-PARC E57 (SIDDHARTA-2).

In the present work we investigate the level shifts and
decay widths of kaonic deuterium by performing a full three-body calculation, aiming at 
a more stringent constraint for the $I=1$ component of the $\bar{K}N$ interaction.
We use a modern $\bar{K}N$ interaction, the Kyoto $\bar{K}N$ potential~\cite{mh} that has been
constructed based on chiral SU(3) effective field theory.
Thanks to systematic improvements with inclusion of higher order terms, 
this potential reproduces very well all available low-energy $\bar{K}N$ empirical data ($K^-p$ total cross sections,
branching ratios at $K^-p$ threshold and the SIDDHARTA data) with an accuracy of $\chi^2/\text{d.o.f.}\simeq
1$~\cite{ihw1,ihw2}. 

There are several theoretical studies of kaonic deuterium
as an approximate $K^-d$ two-body problem~\cite{kdshe,kdiva,kdbar}.
The relationship between kaonic deuterium observables and $\bar{K}N$ two-body scattering lengths has been discussed in Refs.~\cite{Meissner:2006gx,Gal:2007,Doring:2011xc}. Three-body Faddeev calculations of kaonic deuterium also have their own history - see for example Ref.~\cite{Bahaoui:2003}.
Recently, advanced Faddeev calculations with separable potentials
constrained by the SIDDHARTA kaonic hydrogen data have been performed~\cite{kdrev,kddol}. The 
$1S$ $K^-d$ atomic state was evaluated assuming isospin symmetry
for the $\bar{K}$ and nucleon doublets. 

Experimentally the $1S$ level shift is determined
from the transition between $2P$ and $1S$ states.
It is thus important to calculate the precise transition energy between these two states.
We perform highly accurate three-body calculations for
the $1S$, $2P$, and $2S$ energy levels using physical masses of all particles involved,
and examine how those levels are affected by the strong $\bar{K}N$ interaction.
Finally, we quantify the accuracy requirement
for a measurement of the kaonic deuterium level shift and width aimed at
improving constraints on the $I=1$ components in the $\bar{K}N$ interaction.

The following section discusses first the properties of the $\bar{K}N$ interaction
by examining the kaonic hydrogen two-body system. Details of the three-body calculation of kaonic
deuterium are explained in Sec.~\ref{threebody.sec}.
Section~\ref{results.sec} presents numerical results of the full 
three-body calculations.
In Sec.~\ref{spect.sec}, 
the energy spectrum of kaonic deuterium
is presented. The impact of strong-interaction effects
on the spectrum is discussed by comparing spectra
with and without the $\bar{K}N$ interaction.  
In Sec.~\ref{sens.sec} the
sensitivity with respect to the $I=1$ component of the $\bar{K}$$N$ interaction 
is tested by changing selectively its strength within presently existing uncertainties.
In Sec.~\ref{deser.sec} we compare our results to the improved Deser formula, a frequently used
approximation to evaluate the $1S$ level shift.
A summary is given in Sec.~\ref{summary.sec}.

\section{$\bar{K}$$N$ INTERACTION and Kaonic hydrogen}
\label{khydro.sec}

In the present work we employ the recently developed, complex and energy dependent Kyoto $\bar{K}N$ potential~\cite{mh} as the basic antikaon-nucleon interaction. This potential reproduces the scattering
amplitudes that were calculated previously based on chiral SU(3) coupled-channels dynamics~\cite{ihw1,ihw2}. The potential in its original form is written in the isospin basis, from which the particle basis potential is constructed as
\begin{align}
 \hat{V}^{\bar{K}N}_{ij}&=\frac{1}{2}\left[\hat{V}^{\bar{K}N(I=0)}_{ij}+\hat{V}^{\bar{K}N(I=1)}_{ij}\right]\nonumber\\
 &\quad -\frac{1}{2}\left[\hat{V}^{\bar{K}N(I=0)}_{ij}-\hat{V}^{\bar{K}N(I=1)}_{ij}\right]P_{\tau}^{ij},
\end{align}
where the Heisenberg operator $P_{\tau}^{ij}=(1+\tau_i\cdot\tau_j)/2$
exchanges the $i$- and $j$-th particles in the isospin wave function. In the particle basis which is going to be used in the following, the exchange operator acts as $P_{\tau}^{ij}|K^-n\rangle=|K^-n\rangle$, $P_{\tau}^{ij}|K^-p\rangle=-|\bar{K}^0n\rangle$ and
$P_{\tau}^{ij}|\bar{K}^0n\rangle=-|K^-p\rangle$. 
The charge exchange channel-coupling between $K^{-}p$ and $\bar{K}^{0}n$ occurs through the isospin dependence of the $\bar{K}N$ interaction.

Effects of the decay processes into $\pi\Sigma$ and $\pi\Lambda$ channels are encoded in the imaginary part of the $\bar{K}N$ potential.
The strength of the interaction depends on the energy $E_{\bar{K}N}$, to be treated self-consistently in the Schr\"odinger equation in order to reproduce the coupled-channels scattering amplitudes.

Before proceeding to the calculation of kaonic deuterium, it is mandatory to check the applicability of the Kyoto  $\bar{K}N$ potential for the present study. We recall that the fitting to the kaonic hydrogen data in Refs.~\cite{ihw1,ihw2} was performed making use of the improved Deser formula~\cite{deser}, whereas in the present work we solve the Schr\"odinger equation to evaluate the level shift and width. Moreover, the threshold energy difference between the $K^{-}p$ and $\bar{K^{0}}n$ channels is about 5 MeV and should properly be taken into account in the level shift calculation. While the amplitudes in Refs.~\cite{ihw1,ihw2} were calculated using physical hadron masses, the Kyoto $\bar{K}N$ potential of Ref.~\cite{mh} has originally been formulated in the isospin basis with isospin averaged masses, and so this potential is converted to the particle basis with physical masses for the application reported in the present work. 

As a first test the energy of kaonic hydrogen is computed by solving
the coupled-channels Schr\"odinger equation
 \begin{align}
&
   \begin{pmatrix}
    \hat{T}+\hat{V}^{\bar{K}N}+\hat{V}^{\rm C} & \hat{V}^{\bar{K}N}\\
     \hat{V}^{\bar{K}N}& \hat{T}+\hat{V}^{\bar{K}N}+\Delta m
   \end{pmatrix}
   \begin{pmatrix}
    \left|K^-p\right>\\
    \left|\bar{K}^0n\right>
   \end{pmatrix}\notag\\
   &=   E
   \begin{pmatrix}
    \left|K^-p\right>\\
    \left|\bar{K}^0n\right>
   \end{pmatrix}     
 \end{align}
 with the kinetic energy $\hat{T}$,
 the Coulomb interaction $\hat{V}^{\rm C}$
and the Kyoto $\bar{K}N$ potential in the particle basis,
using the following physical masses:
$M_{p}=938.272$~MeV, $M_{n}=939.565$~MeV, $M_{K^{-}}=493.677$~MeV, and
$M_{\bar{K}^{0}}=497.648$~MeV for proton, neutron and antikaons, respectively~\cite{Olive:2016xmw}.
  $\Delta m$ denotes the mass difference of the $K^-p$
  and $\bar{K}^0n$ channels.
  The kaonic hydrogen wave function is expressed as a superposition of a large number of 
  square-integrable Gaussian basis functions reaching out to far distances.	 
  In the Coulomb-bound state the $K^-p$ channel is closed and
  the $\bar{K}^0n$ channel with its higher physical mass is closed as well,
  and so the relevant matrix elements are accurately determined.

The results for the $K^-$-hydrogen shift and width are listed in Table~\ref{k-p.tab}. As shown in the first line of this table, the self-consistent solution of the Schr\"odinger equation using physical masses reproduces the experimental SIDDHARTA result~\cite{sidd,sidd2} within its uncertainties. The Kyoto $\bar{K}N$ potential in the particle basis thus proves to be a valid input even though the original construction of the potential was not optimized for this purpose. On the other hand, when calculating kaonic hydrogen with isospin-averaged masses of the antikaon and nucleon doublets, we obtain the result shown in the second (``Isospin'') row of Table~\ref{k-p.tab}. One observes a quantitative change of the energy shift by more than 100 eV, exceeding by far the uncertainty of the measurement~\cite{sidd,sidd2}. While it is common practice in strong-interaction calculations to assume that isospin breaking effects are not very significant, these effects can be kinematically enhanced in near-threshold observables. 
   To elucidate the difference, we show in Table~\ref{slength.tab} the $\bar{K}N$ scattering lengths calculated with physical masses and with isospin-averaged masses. The isospin averaging implies an upward shift of the $K^{-}p$ threshold by 2.6 MeV from its physical location. As a consequence the real part of the $K^{-}p$ scattering length $a_{K^{-}p}$ is reduced in magnitude by 0.26 fm (i.e. by about 40 \%). The more detailed discussion of the resulting kaonic hydrogen energy shift and width follows in Section~\ref{deser.sec} featuring the improved Deser formula.
Hence it is obvious that precise physical masses must be used in the level shift computation.

Next we examine the effect of the energy dependence of the Kyoto $\bar{K}N$ potential. This energy dependence is essential in determining the binding energies (several tens of MeV) of $\bar{K}$-nuclear systems with few to several nucleons~\cite{Ohnishi17}. However, the atomic states are located in the near neighborhood of the threshold. Their binding energies are as small as a few keV. To study the effect of the energy dependence, we perform the same calculation as previously described, but setting $E_{\bar{K}N}=0$ in the potential. As shown in the third row of Table~\ref{k-p.tab}, the self-consistent and fixed $E_{\bar{K}N}=0$ results turn out to be numerically identical. Therefore, in the level shift calculation of the atomic states, the energy dependence of the $\bar{K}N$ potential can be safely neglected, and this is how we shall proceed hereafter, setting $E_{\bar{K}N}=0$ throughout.

\begin{table}[tbp]
\caption{Level shifts and decay widths
    of the $1S$ atomic state of the kaonic hydrogen
    with physical masses and with isospin averaged masses.
    Results by setting $E_{\bar{K}N}=0$ in the $\bar{K}N$ interaction are also shown.}
\begin{ruledtabular}
\begin{tabular}{cccccc}
   Mass& $E$-dep. &$\Delta$E (eV) && $\Gamma$ (eV) \\ \hline
   Physical & Self consistent& 283 && 607 \\
   Isospin  & Self consistent & 163 && 574 \\   
   Physical &$E_{\bar{K}N}=0$& 283 && 607 \\ 
\hline
   Expt.~\cite{sidd,sidd2} & & 283 $\pm$ 36 $\pm$ 6 && 541 $\pm$ 89 $\pm$22 \\
\end{tabular}
 \end{ruledtabular}
\label{k-p.tab}
\end{table}%

\begin{table}[tbp]
\caption{$\bar{K}N$ scattering lengths with physical masses and with isospin averaged masses.}
\begin{tabular}{cccccc}
\hline\hline
   Mass & $a_{K^{-}p}$ (fm) & $a_{K^-p\text{-}\bar{K}^0n}$ (fm) & $a_{\bar{K}^{0}n}$ (fm) & $a_{K^{-}n}$ (fm) \\ \hline
   Physical & $-0.66+i0.89 $ & $-0.85+i0.26 $ & $-0.40+i1.03 $ & $0.58+i0.78 $ \\
   Isospin  & $-0.40+i0.81 $ & $-0.99+i0.04 $ & $-0.40+i0.81 $ & $0.58+i0.77 $  \\
\hline\hline
\end{tabular}
\label{slength.tab}
\end{table}%

\section{Three-body approach to kaonic deuterium}
\label{threebody.sec}

\subsection{Three-body Hamiltonian}

We start from the following three-body Hamiltonian for kaonic deuterium:
\begin{equation}
\hat{H}=\sum_{i=1}^3 \hat{T_{i}} 
-\hat{T}_{\rm cm}+\hat{V}_{23}^{NN}+
\sum_{i=2}^3 (\hat{V}_{1i}^{\bar{K}N}+\hat{V}_{1i}^{EM}),
\end{equation}
where $\hat{T_{i}}$ denotes the kinetic energy of the $i$-th particle
($i=1$ for an antikaon, and $i=2$, 3 for two nucleons),
including physical masses of $p$, $n$, $K^-$, and $\bar{K}^0$.
The center-of-mass kinetic energy, $\hat{T}_{\rm cm}$, is properly subtracted.

We use the Minnesota potential~\cite{nn} as the $NN$
interaction, $\hat{V}^{NN}$. This potential is technically convenient for three-body computations. It operates with a central force only but reproduces quantitatively the binding energy and radius of the deuteron. In fact what matters primarily in the kaonic deuterium calculation is a  deuteron density distribution, $\rho_d(r)$. We checked that $r^2\rho_d(r)$ deduced from the Minnesota potential agrees perfectly and quantitatively with the radial density profile generated by realistic $NN$ interactions such as the CD-Bonn potential~\cite{Machleidt:2001}. 

For the antikaon-nucleon interaction, $\hat{V}^{\bar{K}N}(E)$,
we employ the Kyoto $\bar{K}N$  potential~\cite{mh}. As just pointed out, the choice of the two-body antikaon-nucleon energy at threshold, $E \equiv E_{\bar{K}N} = 0$, is justified for kaonic hydrogen. For kaonic deuterium this issue requires further discussion. The energy of the $\bar{K}N$ two-body subsystem within the $K^- d$ three-body system is not a well-defined concept. Different prescriptions \cite{dote1,dote2,barn,Ohnishi17} are available to take into account the motion of the bound nucleons while they interact with the antikaon. In the present work we use the prescription of Refs.~\cite{dote1,dote2,Ohnishi17} where $E_{\bar{K}N}$ is proportional to the kaon binding energy. This amounts to setting $E_{\bar{K}N} = 0$ in the two-body potential  $\hat{V}^{\bar{K}N}$ also for kaonic deuterium, the choice we take as our default input in the following three-body calculations. 
Leading corrections to this minimal choice are discussed in Appendix~\ref{sec:Appendix} and numerically estimated using the resummed Deser formula in Section~\ref{deser.sec}.
%Leading corrections to this minimal choice can roughly be estimated using the improved and resummed Deser formula described in Section IV.C. Here we briefly anticipate the outcome of such an estimate: the changes of $\bar{K}N$ scattering lengths, induced by correcting the energy dependence in $\hat{V}^{\bar{K}N}(E)$ for deuteron binding, tend to increase the width of kaonic deuterium by about 10\%, while the corresponding energy shift changes only marginally. 

The 
electromagnetic (Coulomb) interaction is denoted by $\hat{V}^{EM}$.
The effect of higher order QED corrections 
will be discussed in Sec.~\ref{deser.sec}.
The explicit three-body coupled-channels equation is written as
 \begin{align}
&
   \begin{pmatrix}
    \hat{H}_{K^-pn}& \hat{V}_{12}^{\bar{K}N}+\hat{V}_{13}^{\bar{K}N}\\
    \hat{V}_{12}^{\bar{K}N}+\hat{V}_{13}^{\bar{K}N}&\hat{H}_{\bar{K}^0nn}     
   \end{pmatrix}
   \begin{pmatrix}
    \left|K^-pn\right>\\
    \left|\bar{K}^0nn\right>
   \end{pmatrix}=   E
   \begin{pmatrix}
    \left|K^-pn\right>\\
    \left|\bar{K}^0nn\right>
   \end{pmatrix}     
 \end{align}
 with
 \begin{align}
   \hat{H}_{K^-pn}&=\sum_{i=1}^3\hat{T}_i-\hat{T}_{\rm cm}+\hat{V}_{23}^{NN}+\sum_{i=2}^3(\hat{V}_{1i}^{\bar{K}N}+\hat{V}_{1i}^{\rm EM}), \\
   \hat{H}_{\bar{K}^0nn}&=\sum_{i=1}^3\hat{T}_i-\hat{T}_{\rm cm}+\hat{V}_{23}^{NN}+\sum_{i=2}^3\hat{V}_{1i}^{\bar{K}N}+\Delta M
   \end{align}
 with $\Delta M$ denoting
 the mass difference of the $K^-pn$ and $\bar{K}^0nn$ channels.
 In the following subsection 
we describe how the coupled-channels three-body equation is solved in practice.

\subsection{Basis functions}

The three-body Schr\"{o}dinger equation is solved using
a variational method with basis expansion.
The generic basis function is  expressed as
\begin{equation}
\label{eq:wave}
\Phi=\mathcal{A}[\psi^{\rm (space)}\otimes\psi^{\rm (spin)}\otimes\psi^{\rm (isospin)}],
\end{equation}
where $\mathcal{A}$ is the antisymmetrizer for two nucleons.

Since the Hamiltonian considered in this paper does not change
the total orbital angular momentum, $L$, and the total spin, $S$,
of the particles, we can introduce 
an $L=0$ and $S=1$ state with isospin-$3$-component $M_{T}=-\frac{1}{2}$ 
as a basis to describe kaonic deuterium.
The spin wave function $S=1$ is given explicitly as
\begin{equation}
\psi^{\rm (spin)}=\frac{1}{\sqrt{2}}\bigl(|\uparrow\downarrow\rangle+|\downarrow\uparrow\rangle\bigr) ,
\end{equation}
where the first (second) arrow indicates the spin of the nucleon with index $i=2$ ($i=3$).
The isospin part, $\psi^{\rm (isospin)}$, of the wave function written in the particle basis
includes the following two channels:
\begin{equation}
|K^{-}pn\rangle=|\downarrow\uparrow\downarrow\rangle,\hspace{5mm}
 |\bar{K}^{0}nn\rangle=|\uparrow\downarrow\downarrow\rangle.
 \label{isospin_wave}
\end{equation}

For the radial part of the wave function we use correlated Gaussian (CG)
basis functions~\cite{cgpaper,cgtext}.
This method is sufficiently flexible so that it enables us
to describe both short- and long-range behaviors of the
wave function accurately, a necessary condition when dealing with systems such as kaonic deuterium in which the very different distance scales characteristic of Coulomb and strong interactions must be treated simultaneously. (See recent reviews~\cite{cgrev,clust} for many applications of the CG method.)

Let $\bm{x}$ denote a two-dimensional column vector
whose $i$-th element is
a usual 3-dimensional coordinate vector, ${\bm{x}}_{i}$.
The spatial part of the wave function in Eq.~(\ref{eq:wave})
is written in the form~\cite{suzuki08}
\begin{equation}
\label{eq: radwave}
F_{LM_{L}}(u,v,A,\bm{x})=\exp(-\frac{1}{2}\tilde{\bm{x}}A\bm{x})
\left[{\mathcal{Y}}_{L_1}(\tilde{u}\bm{x}){\mathcal{Y}}_{L_2}(\tilde{v}\bm{x})
 \right]_{LM_L},
\end{equation}
with solid spherical harmonics
\begin{equation}
{\mathcal{Y}}_{lm}(\bm{r})=r^{l}Y_{lm}(\hat{\bm{r}}).
\end{equation}
In Eq.~(\ref{eq: radwave}), $A$ is a $2\times2$ positive-definite symmetric matrix,
and a tilde stands for the transposed matrix. The product
$\tilde{\bm{x}}A\bm{x}$ is a short-hand notation
for $A_{11}x_{1}^{2}+A_{22}x_{2}^{2}+2A_{12}{\bm{x}}_{1}\cdot{\bm{x}}_{2}$.
The off-diagonal element, $A_{12}$, induces correlations
between the coordinates $\bm{x}_1$ and $\bm{x}_2$.
The global vector (GV),
$\tilde{u}\bm{x}=u_1{\bm{x}}_{1}+u_{2}{\bm{x}}_{2}$
describes rotational motion of the system,
with $u$ and $v$ being two-dimensional column vectors which specify the rotation axes. 

One of the advantages of the combined CG+GV method is that
its functional form does not change under
linear coordinate transformations.
Suppose that the matrix $A$ and the vectors
$u$ and $v$ are defined in the $\bm{x}$ coordinate set.
Defining a transformation matrix $T$
as $\bm{y}=T\bm{x}$, we can work equivalently with the $\bm{y}$ coordinate set, simply
replacing $A$, $u$ and $v$ by $\tilde{T}AT$, $\tilde{T}u$ and $\tilde{T}v$, respectively.

Consider two sets of Jacobi coordinates:
\begin{align}
{\bm{x}}_{1}&={\bm{r}}_{2} - {\bm{r}}_{3},\notag \\
{\bm{x}}_{2}&=\frac{m_{2}}{m_{2}+m_{3}}{\bm{r}}_{2} + \frac{m_{3}}{m_{2}+m_{3}}{\bm{r}}_{3}- {\bm{r}}_{1},
\label{jacobi1}
\end{align}
and
\begin{align}
{\bm{y}}_{1}&={\bm{r}}_{1} - {\bm{r}}_{2},\notag\\
{\bm{y}}_{2}&=\frac{m_{1}}{m_{1}+m_{2}}{\bm{r}}_{1} + \frac{m_{2}}{m_{1}+m_{2}}{\bm{r}}_{2}- {\bm{r}}_{3},
\label{jacobi2}
\end{align}
where ${\bm{r}}_{i}$ stands for the single-particle coordinate of the $i$-th particle. Both of these sets are equally suitable for the three-body calculation.
However, when dealing with the channel coupling between systems of different mass, the coordinates
$\bm{x}_2$ and $\bm{y}_2$ are not common to the
$K^{-}pn$ and $\bar{K^{0}}nn$ channels. We therefore use the following integral coordinates
which do not depend on any particle masses:
\begin{align}
{\bm{z}}_{1}&={\bm{r}}_{1} - {\bm{r}}_{2},\notag \\
{\bm{z}}_{2}&={\bm{r}}_{2} - {\bm{r}}_{3},
\end{align}
for evaluating the off-diagonal matrix element
between the $K^{-}pn$ and $\bar{K^{0}}nn$ channels.

\subsection{Energy convergence}

In this work short-range 
strong interactions as well as the long-range Coulomb interaction have to be treated
simultaneously with high precision. To extract the detailed effects of the $\bar{K}N$ interaction from
the spectrum of kaonic deuterium,
we need to calculate the binding energy with an accuracy of a few eV.
This is a computational challenge that demands great care.
In this subsection, we discuss how to meet this challenge of calculating wave functions with the required precision. 

The wave function is expanded in a large set of
basis functions, Eq.~(\ref{eq:wave}),
and the generalized eigenvalue problem 
\begin{align}
\sum_{j=1}^K(H_{ij}-EB_{ij})C_j=0,
\end{align}
is solved to determine the
coefficients $C_i$ and eigenenergy $E$,
with the Hamiltonian matrix $H_{ij}=\left<\Phi_i|H|\Phi_j\right>$ and the overlap matrix
$B_{ij}=\left<\Phi_i|\Phi_j\right>$.
Here $K$ is the number of basis functions.
To achieve energy convergence for the kaonic atom,
it turns out that we need to include basis functions
reaching over distance scales from one tenth to several hundreds of fm.
Given the large number of non-orthogonal basis functions,
we cannot solve the generalized eigenvalue problem
due to round-off errors in the double precision computation~\cite{hiyama12}.
To avoid this problem, we reconstruct a
new orthonormal basis set from the prepared basis functions
by diagonalizing the overlap matrix $B_{ij} $:
\begin{equation}
\phi_{\mu}=\frac{1}{\sqrt{\mu}}\sum_{i=1}^{K}c_{i}^{(\mu)}\Phi_i.
\end{equation}
The number of new basis functions $\{\phi_{\mu}\}$ is again $K$,
and each function is labeled by its eigenvalue $\mu$. 
The Hamiltonian is then diagonalized with this set of basis functions, 
omitting those which give very small $\mu$.
If a whole set of basis functions emerges with very small $\mu$,
we discard this set altogether
and try another one.
In practice a cutoff parameter is introduced,
defined by the ratio of minimum to maximum eigenvalues $\mu$
as $\lambda_{\rm cut}=\mu_{\rm max}/\mu_{\rm min}$. Basis functions with $\mu<\mu_{\rm min}$
are discarded. The cutoff parameter is taken as large as possible
within significant digits of the double precision computation.

To generate the elements of the matrix $A$
(the variational parameters), we use
a geometric progression~\cite{geo}
for diagonal matrix elements of $A$
with the $\bm{x}$ coordinates defined in Eq.~(\ref{jacobi1}).
For the global vectors,
we simply take $\tilde{u}=(1,0)$ and $\tilde{v}=(0,1)$ to
define an angular momentum for each coordinate.
Intermediate angular momenta up to $L_1+L_2\leq 4$ are taken into account.

For the diagonal elements of the matrices $A$, $u$, and $v$, the variational procedures 
can actually be optimized by suitably combining a representation using the coordinates $\bm{x}$
of Eq.~(\ref{jacobi1}) with the equivalent representation in the so-called rearrangement channel, using the 
coordinates $\bm{y}$ of Eq.~(\ref{jacobi2}). 
The evaluation of the Hamiltonian matrix elements is then
performed in $\bm{x}$ coordinates applying the  
 transformations $A\to \tilde{T}AT$, $u\to \tilde{T}u$ and $v\to \tilde{T}v$ where appropriate.

With one-by-one inclusion of those channels just mentioned,
several sets of variational parameters are prepared
covering distance scales from 0.1 fm to 300-1000 fm,
in search for the lowest energy.
We need more than 30 Gaussian basis functions for each coordinate
to achieve energy convergence within a few eV.
After a careful examination of the energy convergence
by introducing the cutoff parameter $\lambda_{\rm cut}$,
the total number of basis functions $K$ is 4096 and 8192
for the $S$ and $P$ states, respectively.

Table~\ref{conv.tab} shows the cutoff dependence of the real part of
the energy of the kaonic deuterium $1S$ state measured from the three-body break-up threshold.
$N$ denotes the number of basis functions that actually appear in the diagonalization.
The number of primary basis functions, $K=4096$,
is reduced with decreasing $\lambda_{\rm cut}$.
It turns out that we cannot diagonalize the Hamiltonian for 
$\lambda_{\rm cut}\gtrsim 10^{23}$ due to round-off errors in
the double precision calculations.
Finally we reach convergence within eV accuracy
for $\lambda_{\rm cut} \gtrsim 10^{20}$,
in which case the number of basis functions
becomes approximately half of the number of primary basis functions.
For the $2P$ state, we take $\lambda_{\rm cut} \gtrsim 10^{28}$, 
and $N\gtrsim 3508$ basis functions are actually needed in the diagonalization.

\begin{table}[tbp]
\caption{Cutoff parameter $\lambda_{\rm cut}$, number of basis functions $N$, and the real part of the energy
    of the $1S$ state of kaonic deuterium.}
		\begin{ruledtabular}
\begin{tabular}{ccc}
$\log_{10}\lambda_{\rm cut}$& $N$ & Re[$E$] (MeV)\\
\hline
16&	1677&	$-$2.211689436\\
17&	2194&	$-$2.211722964\\
18&	2377&	$-$2.211732072\\
19&	2511&	$-$2.211735493\\
20&	2621&	$-$2.211737242\\
21&	2721&	$-$2.211737609\\
22&	2806&	$-$2.211737677\\
23&	2879&	$-$2.211737682\\ 
\end{tabular}
\end{ruledtabular}
\label{conv.tab}
\end{table}%

\section{Results and discussion}
\label{results.sec}

\subsection{Spectrum and level shifts}
\label{spect.sec}
\begin{table*}[tbp]
\caption{Energy spectrum of kaonic deuterium.
 Three- and two- body calculations with Coulomb interaction only
  (omitting the strong $\bar{K}N$ interaction) are listed in the first three rows.
Energy levels resulting from the three-body calculation are measured relative to
the calculated $K^-d$ threshold.
For the $K^-d$ two-body calculations
the deuteron mass $M_d=1875.613$\,MeV has been used~\cite{Olive:2016xmw}.}
		\begin{ruledtabular}
\begin{tabular}{cccccc} 
                &$E_{1S}$(keV) & $E_{2P}$(keV) & $E_{2S}$(keV) \\ 
\hline
Coulomb       & $-$10.398 & $-$2.602& $-$2.600\\
Uniform charge (2-body)   & $-$10.401 & $-$2.602& $-$2.601\\
Point charge (2-body)     & $-$10.406 & $-$2.602& $-$2.602\\ 
\hline
Coulomb$+\bar{K}N$&$-$9.736$-i\,$0.508& $-$2.602$-i\,$0.000&$-$2.517$-i\,$0.067\\ 
\end{tabular}
        \end{ruledtabular}
\label{fullresult}
\end{table*}%

Table~\ref{fullresult} lists
binding energies, measured from the $K^-d$ threshold,
and decay widths of kaonic deuterium. 
The three-body calculation with Coulomb
interaction only is shifted slightly from the energy levels
produced in the $K^-d$ two-body calculations
with point charge, by 8 eV and 1 eV for
the $1S$ and $2S$ states, respectively. The $2P$ energy
remains unchanged in the three-body calculation because
the $P$-wave function around the origin is suppressed by
the centrifugal barrier.
This behavior is consistent with the $K^-d$ two-body 
estimate of the energy shift assuming a uniform charge distribution
as listed in the table.

With inclusion of the $\bar{K}N$ interaction
the $1S$ state is shifted by $\sim 670$ eV
from the $K^-d$ Coulomb (point charge) $1S$ level. 
The level shift and width of the $2S$ level are an order of magnitude smaller than
those of the $1S$ state because the $2S$ wave function has a smaller
amplitude around the origin than the one of the $1S$ state.
The $2P$ energy remains unchanged and its
decay width is found to be less than 1 eV;
the $\bar{K}N$ interaction has virtually no effect on the $2P$ state
of kaonic deuterium because of the presence of the centrifugal barrier.
We can therefore safely extract the $1S$ level shift from the $2P \to 1S$
transition energy.
In summary, the $1S$ level shift and decay width resulting from the full three-body calculation are predicted as:  
\begin{equation}
\Delta E-i\frac{\Gamma}{2}= (670-i\,508)\text{\,eV},
\end{equation}
namely, $(\Delta E,\Gamma)=(670, 1016)$ eV using the Kyoto $\bar{K}N$ potential.
These values are roughly consistent with those found in a recent Faddeev calculation~\cite{kdrev},
although the basic interactions used in that approach are different from ours.

For comparison, a full three-body computation of the level shift and width has also been performed using isospin-averaged meson and baryon masses, with the result $\Delta E-i\Gamma/2=(672-i\,509)$\,eV. The small deviation, by just a few eV, from the corresponding calculation using physical masses is of some interest here, as this is in unexpected contrast to the relatively large isospin-breaking effects seen in kaonic hydrogen. Some insight into the origin of this difference can be gained by a closer look into the multiple scattering series and the improved Deser formula which relates the level shift and width to the pertinent scattering lengths, see subsection \ref{deser.sec}.

Up to this point the determination of the width $\Gamma$ incorporates the decay channels $\bar{K}N\rightarrow\pi Y$, where $Y$ stands for $\Lambda$ and $\Sigma$ hyperons. The question arises about possible additional contributions to the width from antikaon absorption on two nucleons, with the coupled $K^-pn$ and $\bar{K}^0 nn$ channels decaying into $\Lambda n + \Sigma^0 n + \Sigma^- p$. Early measurements at Brookhaven with $K^-$ stopped on liquid deuterium in the BNL bubble chamber~\cite{Veirs:1970fs} demonstrated that these processes are strongly suppressed as compared to the leading single-nucleon channels, $\bar{K}N\rightarrow \pi Y$. The ratio of two-nucleon absorption reactions to the single-nucleon processes was found to be as small as $(1.2\pm 0.1)\%$~\cite{Veirs:1970fs}. Taking this value for orientation, the kaonic deuterium $1S$ width would increase through two-nucleon absorption by only about $10$ eV, a correction that can be safely neglected within an uncertainty range of approximately $10\,\%$ assigned to the calculated width of about a keV. The smallness of the two-body absorptive width can be understood as follows. Kinematical conditions for the $\bar{K}NN\rightarrow YN$ process require a large momentum transfer of order 1 GeV/c to be provided by the initial deuteron wave function at short distances. The probability for this to take place in a weakly bound, dilute system like the deuteron is small. Similar considerations hold, for example, in the analysis of the $^3$He$(K^-,\Lambda p)n$ reaction~\cite{jparc2}. Background simulations performed for this experiment pointed out that two-nucleon absorption is strongly suppressed in the vicinity of the $K^-pp$ threshold, whereas three-nucleon reactions dominate.

\subsection{Constraining the $I=1$ component of $\bar{K}N$ interaction}
\label{sens.sec}

To quantify the sensitivity of the kaonic deuterium level shift 
with respect to the $I=1$ component of the $\bar{K}N$ interaction,
we vary its strength within the uncertainties of 
the SIDDHARTA kaonic hydrogen measurement~\cite{sidd,sidd2}.
This uncertainty range can be simulated by simply multiplying a constant, $\beta$, to
the real part of the $I=1$ component of the $\bar{K}N$potential.
Within the SIDDHARTA constraint~\cite{sidd,sidd2}, the control parameter
$\beta$ can range from $-$0.17 to 1.08. Evidently this constraint is quite weak:
even $\beta=0$, i.e. a vanishing real part of the $I=1$ $\bar{K}N$ potential, would still be acceptable.
Theoretical considerations based on chiral $SU(3)$ dynamics would exclude such a possibility, but it cannot be ruled out by just looking at the SIDDHARTA data.

Table~\ref{jyusui} lists the results of the two- and three-body calculations
performed with limiting values of $\beta$ compared to the standard case, $\beta = 1$.
It is interesting to observe that the sensitivity with respect to the $I=1$ $\bar{K}N$ interaction strength
shows different patterns for $\Delta E$ and $\Gamma$ in kaonic hydrogen as compared to
kaonic deuterium.
In the $K^-p$ system, a variation of  $\beta$ within its upper and lower limits changes 
$\Delta E$ by less than 10\%, whereas $\Gamma$ changes by more than 30\%. On the other hand, the same variation of $\beta$ in the $K^-pn$ system induces a change $\Delta E$ by 170\,eV while $\Gamma$ remains stable around 1 keV. 

One concludes that an accuracy of about 25\% in a measurement of the 1S shift in kaonic deuterium would already improve the determination of the $I=1$ $\bar{K}N$ interaction
considerably over the kaonic hydrogen result. The 30-60\,eV precision to be expected in the planned experiments~\cite{sidd-2,Iliescu:2016zgx} falls well within that range.

\begin{table}[htb]
  \centering
  \caption{Level shifts and decay widths (in eV) of kaonic hydrogen and deuterium
computed with different $I=1$ strengths of the $\bar{K}N$ interaction.
The experimental level shift data of kaonic hydrogen is 
$(\Delta E,\Gamma)=(283\pm 36 \pm 6$, $541 \pm 89 \pm 22)$ eV~\cite{sidd,sidd2}.}
  \begin{tabular}{cccccc} 
\hline \hline
 &\multicolumn{2}{c}{$K^-p$}&&\multicolumn{2}{c}{$K^-d$}\\
\cline{2-3}\cline{5-6}
 $\beta$~ &~~~$\Delta$E~~~~&$\Gamma$~~~&&~$\Delta$E~~~&$\Gamma$~~ \\ \hline
   1.08       &287 & 648 && 676 & 1020 \\
   1.00       &283 & 607 && 670 & 1016 \\
   $-$0.17    &310 & 430 && 506 & 980 \\ 
   \hline \hline
  \end{tabular}
  \label{jyusui}
\end{table}

\subsection{Improved Deser formulae for kaonic deuterium}
\label{deser.sec}

The improved Deser formula~\cite{deser,Meissner:2006gx},
derived from non-relativistic effective field theory (EFT),
is frequently used in the investigation of strong-interaction effects in hadronic atoms.
The $1S$ level shift $\Delta E$ and width $\Gamma$ of a kaonic atom can be estimated by the relation~\cite{deser,Meissner:2006gx}:
\begin{align}
 \Delta E - \frac{i\Gamma}{2}=-2\mu^2\alpha^3a[1-2\mu\alpha(\ln
 \alpha-1)a],\label{imp_deser}
\end{align}
where $\mu$ is the kaon-nucleus reduced mass, $\alpha$ is the fine
structure constant and $a$ is the $K^-$-nucleus scattering length.
The logarithmically enhanced correction term can be resummed to all orders~\cite{Baru:2009tx}, providing a ``double-improved" Deser formula: 
\begin{align}
 \Delta E - \frac{i\Gamma}{2}=-\frac{2\mu^2\alpha^3a}{1+2\mu\alpha(\ln
 \alpha-1)a}.\label{imp_deser2}
\end{align}
In this section we compare our full three-body calculation results with the
results obtained from Eqs.~(\ref{imp_deser}) and (\ref{imp_deser2}). But let us first examine the shift and width of kaonic hydrogen in this context. The $K^-p$ scattering length obtained by solving the two-body Schr\"odinger equation with the Kyoto $\bar{K}N$ potential is 
  shown in Table~\ref{slength.tab}.
Using Eqs.~(\ref{imp_deser}) and (\ref{imp_deser2}) one finds the results shown in Table~\ref{tbl:Kpresult}. It is evident that the improved Deser formula works reasonably well for kaonic hydrogen, and the resummed version indeed improves the accuracy further. 

\begin{table}[tbp]
\caption{Level shift and width of kaonic hydrogen obtained by solving the Schr\"odinger equation with the Kyoto $\bar{K}N$ potential, and by using the improved Deser formula and its resummed version.} 
\begin{center}
\begin{ruledtabular}
\begin{tabular}{lcc}
   & $\Delta E$ (eV) & $\Gamma$ (eV)  \\
   \hline
   Full Schr\"odinger equation & 283 & 607  \\
   Improved Deser formula~\eqref{imp_deser} & 293 & 596  \\
   Resummed formula~\eqref{imp_deser2} & 284 & 605 \\
\end{tabular}
\end{ruledtabular}
\end{center}
\label{tbl:Kpresult}
\end{table}%

Estimates of the level shift and width of kaonic deuterium using the Deser formulae require the $K^{-}d$ scattering length $a_{K^{-}d}$ as input. In the fixed center approximation (FCA) for the nucleons,
$a_{K^-d}$ derived from a multiple scattering series is given as~\cite{Kamalov:2000iy,Meissner:2006gx}
\begin{align}
a_{K^-d}&=\frac{\mu_{K^- d}}{m_{K^{-}}}\int d^3\bm{r}\, \rho_d(\bm{r})\,\tilde a_{K^-d}(r),\label{rusetsky}\\
\tilde a_{K^-d}(r)&=\frac{\tilde a_{p}+\tilde a_{n}
+(2\tilde a_{p}\tilde a_{n}-\tilde a_{\rm ex}^2)/r
-2\tilde a_{\rm ex}^2\tilde
a_{n}/r^2}
{1-\tilde a_{p}\tilde a_{n}/r^2+\tilde
a_{\rm ex}^2\tilde a_{n}/r^3},
\label{eq:akd}
\end{align}
with the $K^-$-deuteron reduced mass $\mu_{K^{-}d}$, and $\rho_d(\bm{r})$ is the nucleon density distribution in the deuteron, obtained in the present case using the Minnesota potential. The scattering lengths are defined as 
$\tilde{a}_{p}\equiv\tilde{a}_{K^{-}p}$, $\tilde{a}_{n}\equiv\tilde{a}_{K^{-}n}$ and $\tilde a_{\rm ex}^2\equiv
\tilde a_{K^-p\text{-}\bar{K}^0n}^2/(1+\tilde a_{\bar{K}^0n}/r)$,
and the scattering lengths $\tilde a_{\bar{K}N}$ in the laboratory
frame are given as
$\tilde a_{\bar{K}N}\equiv \frac{m_K}{\mu_{\bar{K}N}}a_{\bar{K}N}$
with the $\bar{K}N$ reduced mass $\mu_{\bar{K}N}$. Using the Kyoto $\bar{K}N$ potential, the resulting 
two-body $\bar{K}N$ scattering lengths 
  are shown in Table~\ref{slength.tab}.
These scattering lengths are defined by the scattering amplitudes at the threshold energy for the diagonal channels and at the average of the threshold energies for the off-diagonal $K^-p$-$\bar{K}^0n$ channel. Their real and imaginary parts agree well with the original amplitudes~\cite{ihw1,ihw2} within their uncertainties. The $K^{-}d$ scattering length is then calculated from Eqs.~(\ref{rusetsky}) and (\ref{eq:akd}) as
\begin{align}
 a_{K^-d}= (-1.42 + i \,1.60)\,\text{fm}.
 \label{eq:aKd}
\end{align}
This result remains unchanged when we adopt a realistic deuteron wavefunction (including the $D$-wave component) generated from the CD-Bonn potential~\cite{Machleidt:2001}. 

Next we apply the improved Deser formulae~(\ref{imp_deser}) and (\ref{imp_deser2}) to kaonic deuterium. The results are summarized in Table~\ref{tbl:Kdresult} together with that from the full three-body calculation. The logarithmic correction term is now increased as $|\mu_{K^{-}d}\,a_{K^{-}d}/(\mu_{K^{-}p}\,a_{K^{-}p})|\sim 1.3$, so the difference between Eqs.~\eqref{imp_deser} and \eqref{imp_deser2} becomes larger than that in kaonic hydrogen. In addition, the deviation from the full three-body calculation is of the order of $\gtrsim$100 eV. 

\begin{table}[tbp]
\caption{Level shift and width of kaonic deuterium obtained by solving the three-body Schr\"odinger equation with the Kyoto $\bar{K}N$ potential, and by using the improved Deser formula and its resummed version.}
\begin{center}
\begin{ruledtabular}
\begin{tabular}{lcc}
   & $\Delta E$ (eV) & $\Gamma$ (eV)  \\
   \hline
   Full Schr\"odinger equation & 670 & 1016  \\
   Improved Deser formula~\eqref{imp_deser} & 910 & 989  \\
   Resummed formula~\eqref{imp_deser2} & 818 & 1188 \\
\end{tabular}
\end{ruledtabular}
\end{center}
\label{tbl:Kdresult}
\end{table}%

Note however that the $K^{-}d$ scattering length in Eq.~\eqref{eq:aKd} is estimated in the FCA limit. Hence it can be different from the exact value. For instance, the importance of recoil corrections, naturally included in the full three-body calculation but neglected in FCA, is discussed  in Refs.~\cite{Baru:2009tx,Mai:2014uma}. In addition, the determination of the precise energy of the two-body $\bar{K}N$ system is subject to some uncertainties.

Another source of small deviations are higher order QED corrections 
such as electron vacuum polarization.
This effect can be included as an effective potential modifying the Coulomb interaction in the form~\cite{greiner}:
\begin{equation*}
V(r)=-\frac{\alpha}{r}\left[ 1+ \frac{2\alpha}{3\pi}\int^{\infty}_{1}due^{-2m_eru} \left( 1+\frac{1}{2u^{2}} \right) \frac{\sqrt{u^{2}-1}}{u^{2}} \right],
\end{equation*}
where $m_e$ is the electron mass. The first term is the ordinary Coulomb potential,
and second term (the Uehling potential)
takes into account the vacuum polarization effect which is found to be small:
The $1S$ level shift and width of the kaonic deuterium including this correction is 
$\Delta E-i\Gamma/2=(670-i\,519)$ eV. 
While the level shift is unchanged, the decay width increases slightly by about 10 eV because 
the Uehling potential is attractive at very short distances.

In summary the improved Deser formulae work well for kaonic hydrogen but estimates based on these formulae appear to be less accurate for kaonic deuterium which does require a three-body treatment beyond fixed nucleons if the aim is to reach a precision at the 10 eV level. 

At this point we can add a comment on the previously mentioned surprising fact that isospin-breaking effects, using physical masses of antikaons and nucleons, are large in kaonic hydrogen but turn out to be small in the full three-body calculation of kaonic deuterium. One can trace this phenomenon by examining the improved Deser formulae together with the multiple scattering relation (\ref{eq:akd}). The prime source of the strong effect in kaonic hydrogen is a substantial change of the real part of the $K^-p$ scattering length 
when using isospin-averaged instead of physical masses. In kaonic deuterium, on the other hand, the whole set of scattering lengths 
  in Table~\ref{slength.tab}
enters Eq.~(\ref{eq:akd}), including $a_{K^-n}$ with its positive real part, so that the leading effect from $a_{K^-p}$ is largely compensated. As a consequence, real parts of $a_{K^-d}$ calculated with physical or isospin-averaged masses now differ only by less than $5\,\%$, and this difference is averaged out further in the full three-body approach beyond fixed-scatterer approximation.
 
Finally we examine possible uncertainties related to the energy dependence of the $\bar{K}N$ potential, $\hat{V}^{\bar{K}N}(E_{\bar{K}N})$. In the present study we have set $E_{\bar{K}N} = 0$ at threshold, following Refs.~\cite{dote1,dote2,Ohnishi17}. The binding of the nucleons in the deuteron may cause a shift of $E_{\bar{K}N}$ towards the subthreshold region. In fact, the prescription in Ref.\,\cite{barn} gives a large negative value for $E_{\bar{K}N}$. Our estimate, derived and discussed in Appendix~\ref{sec:Appendix}, suggests instead a small average shift, $E_{\bar{K}N} =-B_{d}/2\sim -1.1$ MeV, involving the deuteron binding energy $B_{d}$. 
With this value we calculate the level shift and width of kaonic deuterium using the resummed Deser formula~\eqref{imp_deser2} and find $(\Delta E,\Gamma)=(869,1310)\text{ eV}$, compared to $(\Delta E,\Gamma)=(818,1188)\text{ eV}$ with $E_{\bar{K}N}=0$ (see Table~\ref{tbl:Kdresult}).
Thus the changes induced by correcting the energy dependence in $\hat{V}^{\bar{K}N}(E_{\bar{K}N})$ for deuteron binding tend to increase the width of kaonic deuterium by about 10\%, while the corresponding energy shift changes only marginally. 

\section{Conclusions}
\label{summary.sec}

Precise three-body calculations have been performed for the spectrum of
kaonic deuterium and the evaluation of the $1S$ level shift and width.
The $\bar{K}NN$ three-body wave function is expressed by a superposition of a large set of
correlated Gaussian basis functions. In order to describe both
short-range strong interactions and the long-range Coulomb
interaction simultaneously,
a large model space needs to be considered covering all distance scales ranging from 0.1 fm to several hundreds
of fm.

The $\bar{K}N$ strong interaction is treated in terms of a complex potential that accurately reproduces previous results of coupled-channels calculations based on chiral $SU(3)$ dynamics. We have calculated the energy levels of $1S$, $2S$ and $2P$ kaonic deuterium states
and find that the $\bar{K}N$ strong interaction affects only the $S$ states,
inducing energy shifts from the levels characteristic of the pure Coulomb and point charge limit
of the $K^-d$ atomic system. No energy shift
is found for the $2P$ state, so that
the $1S$ level shift can be directly associated with the transition energy from the $2P$ to the $1S$ state. 
The calculated $1S$ level shift of kaonic deuterium is
$\Delta E-i\Gamma/2$=$(670-i\,508)$ eV, corresponding to a $2P \rightarrow 1S$ transition energy of  $7.134$ keV.
Following our previous discussions we assign uncertainies of about 10\% to $\Gamma$ and less than 10\% to $\Delta E$ (not counting the approximately 20\% uncertainties in the empirical SIDDHARTA kaonic hydrogen constraints ). 

In view of upcoming experimental investigations we have also performed a test of the sensitivity of kaonic deuterium observables with respect to the $I=1$ component in the $\bar{K}N$ interaction, by varying selectively
the real part of the $I=1$ $\bar{K}N$ potential strength within the uncertainty limits deduced from the kaonic hydrogen data. 
One can conclude from this test that the $1S$ level shift of kaonic deuterium is indeed expected to provide a significantly improved constraint on the $I=1$ component, as compared to the SIDDHARTA kaonic hydrogen measurement~\cite{sidd,sidd2},
if the deuterium level shift can be determined within $\sim$25\% accuracy (corresponding to $\sim$2\% in the $2P\rightarrow1S$ transition energy). This sets the physics focus on the yet basically unknown $K^-$neutron sector of the $\bar{K}N$ interaction.\\

\acknowledgments

We thank Prof. J. R\'evai and K. Miyahara for valuable communications. 
The numerical calculations were in part performed
on the supercomputer (CRAY XC40) at
the Yukawa Institute for Theoretical Physics (YITP), Kyoto University.
This work is in part supported by the Grants-in-Aid
for Scientific Research
on Innovative Areas from MEXT (Grant No. 2404:24105008),
by JSPS KAKENHI Grant Number JP16K17694 and by the Yukawa International Program for Quark-Hadron Sciences (YIPQS). One of the authors (W. W.) gratefully acknowledges the hospitality of the YITP. He thanks Avraham Gal for instructive exchanges.

\appendix
\section{Two-body $\bar{K}N$ energy and \\deuteron binding correction}
\label{sec:Appendix}
Here we consider the average energy $E_{\bar{K}N}$ of a two-body $\bar{K}N$ subsystem in kaonic deuterium. In general the energy of a two-body subsystem within a three-body bound state is not a well-defined notion. An estimate of $E_{\bar{K}N}$ is nevertheless needed in order to determine the variation of strength of the energy-dependent $\bar{K}N$ potential close to threshold. Prescriptions for such estimates have been discussed in several previous works. 
In the present paper we have argued, following Refs.~\cite{dote1,dote2,Ohnishi17}, that choosing $E_{\bar{K}N}=0$ is a good approximation. Deuteron binding corrections introducing a subthreshold shift, $E_{\bar{K}N}=-B_d/2$ with the deuteron binding energy $B_{d}\sim 2.2$ MeV, imply uncertainties in the kaonic deuterium width of about 10\% and smaller effects on the energy shift. The following considerations are intended to provide a basis for this estimate.
 
% previous prescriptions
A different prescription is given in Ref.\,\cite{barn}, where a considerably larger subthreshold shift,  $E_{\bar{K}N} =-B_{d}/2-M_{N}/(M_{N}+m_{K})\cdot \langle T_{N:N}\rangle/2$, is suggested, involving the pairwise $NN$ kinetic energy $\langle T_{N:N}\rangle\sim 20$ MeV. 
In this case the effective two-body energy turns out to be $E_{\bar{K}N}\sim -7$ MeV, mostly coming from the kinetic energy term. Such a large subthreshold shift in the $\bar{K}N$ potential $V(E_{\bar{K}N})$, when applied in combination with the resummed Deser formula \eqref{imp_deser2}, would produce a massive ($\sim 60\%$) increase of the kaonic deuterium width $\Gamma$ together with an increase of the energy shift $\Delta E$ by about 30\% . 

The expression for $E_{\bar{K}N}$ in Ref.~\cite{barn} is deduced from an {\it ad hoc} ansatz for the average energy of each individual $\bar{K}N$ subsystem within the $\bar{K}$-nuclear many-body system:
\begin{align}
   \sqrt{s_{\rm av}}
   =
   \frac{1}{A}
   \sum_{i=1}^{A}\sqrt{(E_{K}+E_{i})^{2}-(\vec{q}+\vec{p}_{i})^{2}}
   \label{eq:sav}
\end{align}
with the antikaon four-momentum $(E_{K},\vec{q}\,)$, the $i$-th nucleon four-momentum $(E_{i},\vec{p}_{i})$. The number of nucleons is $A=2$ in the present case. 

We argue instead that the starting point for a discussion of $E_{\bar{K}N}$ should be a well-defined quantity, namely the total invariant center-of-mass energy of the 
 $\bar{K}A$ system:
\begin{align}
   \sqrt{s_{\rm tot}}
   =
   \sqrt{
   (E_{K}+E_{A})^{2}-\left(\vec{q}+ \sum_{i=1}^{A}\vec{p}_{i}\right)^{2}
   }
   \label{eq:stot} .
\end{align}
For the kaonic atom case considered here, the Coulomb energy is supposed to be included in $E_{A}$. Eq.\,\eqref{eq:stot} is understood in combination with the constraint of conserved total three-momentum,
\begin{align}
   \vec{P}
   =
   \vec{q}+ \sum_{i=1}^{A}\vec{p}_{i}=\text{const.} 
\end{align}
with $\vec{P}=\vec{0}$ in the rest frame of the kaonic atom.  A decomposition of $\sqrt{s_{\rm tot}}$ into two-body subsystems is then guaranteed to satisfy all kinematic constraints. Note that there is no such systematic link between Eq.\,\eqref{eq:stot} and Eq.\,\eqref{eq:sav}.
% new prescription, Kd

For kaonic deuterium Eq.\,\eqref{eq:stot} becomes:
\begin{align}
   \sqrt{s_{\rm tot}}
   =
   \sqrt{
   (E_{K}+E_{d})^{2}-(\vec{q}+ \vec{p}_{1}+\vec{p}_{2})^{2}
   },
\label{eq:sd}
\end{align}
where $\vec{p}_{1}$ and $\vec{p}_{2}$ now refer to proton and neutron three-momenta within the deuteron. For the purpose of estimating deuteron binding effects, the small Coulomb energy can be dropped. We set, approximately, $E_{K}=m_{K}$ at threshold and $E_{d}=M_{p}+M_{n}-B_{d}$. Three-momentum conservation reads
\begin{align}
   \vec{q}+ \vec{p}_{1}+\vec{p}_{2}
   =
   \vec{0}
\end{align}
in the kaonic deuterium rest frame. In this frame, we have
\begin{align}
   \sqrt{s_{\rm tot}}
   \approx
   m_{K}+M_{p}+M_{n}-B_{d}
   =\sqrt{s_{\rm th}}-B_{d} .
\end{align}
Identifying an average $\bar{K}N$ energy per nucleon as 
\begin{align}
E_{\bar{K}N} = {1\over 2}\left( \sqrt{s_{\rm tot}} - \sqrt{s_{\rm th}} \right)
\end{align}
with $E_{\bar{K}N} = 0$ at threshold, the subthreshold energy shift per nucleon is simply $E_{\bar{K}N}=-B_{d}/2 \simeq -1.1$~MeV. The $\vec{P} = \vec{0}$ constraint implies that
there is no additional strong downward shift from a kinetic energy term, as it would emerge from applying Eq.\,\eqref{eq:sav}.

In order to examine the difference between our approach and the prescription based on Eq.\,\eqref{eq:sav} in more detail, let us expand $ \sqrt{s_{\rm tot}}$ of Eq.\,\eqref{eq:sd} first in an arbitrary frame of reference and write it as a decomposition  into $\bar{K}N$ two-body pieces: 
\begin{eqnarray}
   \sqrt{s_{\rm tot}}&\approx&
   \sqrt{s_{\rm th}}-B_{d} -  {\vec{P}^{\,2}\over 2\sqrt{s_{\rm th}}}~,
\end{eqnarray}
where
\begin{eqnarray}
\vec{P}^{\,2} =
\frac{1}{2}\sum_{i=1,2}\left[(\vec{q}+ \vec{p}_{i})^2 + \vec{p}_{i}^{\,2} + 2\vec{p}_{i}\cdot (\vec{q}+ \vec{p}_{j})\right]
\end{eqnarray}
with $j\neq i$. Note the appearance of cross terms proportional to $\vec{p}_{1}\cdot\vec{p}_{2}$. Such pieces are not present in the expansion of Eq.\,\eqref{eq:sav} which only includes the $(\vec{q}+ \vec{p}_{i})^2$ terms. In the three-body rest frame, $\vec{q}+ \vec{p}_{j} = - \vec{p}_{i}$ so that in the averaged $E_{\bar{K}N}$ the terms of order $\vec{p}^{\,2}$ arrange themselves as:
\begin{eqnarray}
E_{\bar{K}N}=-{B_{d}\over 2}- {1\over 8\sqrt{s_{\rm th}}}\sum_{i=1,2}\langle(\vec{q}+ \vec{p}_{i})^2 - \vec{p}_{i}^{\,2}\rangle~.\nonumber \\
\label{eq:EKbarN}
\end{eqnarray}
Of course the momentum dependent terms vanish altogether as they just reflect the constraint $\vec{P} = \vec{0}$. However, had we kept only the $(\vec{q}+ \vec{p}_{i})^2$ terms and dropped the compensating $- \vec{p}_{i}^{\,2}$ in Eq.\,\eqref{eq:EKbarN}, we would have ended up with a large subthreshold energy shift as in Ref.\,\cite{barn}.

\end{document}